\documentclass[aoas,MSNbibl,nameyear,seceqn,dvips]{arximspdf}
\usepackage{dcolumn}
\usepackage{graphicx}

\doi{10.1214/12-AOAS585} %kopijuoti is PTS
\volume{7}
\issue{1}
\pubyear{2013}
\firstpage{342}
\lastpage{368}

\makeatletter
\newproclaim{definition}{Definition}
\newcolumntype{d}[1]{D{.}{.}{#1}}

\def\real{\mathbb{R}}
\def\bY{\mathbf{Y}}
\def\bX{\mathbf{X}}
\makeatother

\begin{document}
\begin{frontmatter}

\title{Incorporating external information in analyses of clinical
trials with binary outcomes} %\protect\thanksref{T1}}
\runtitle{Incorporating external information in clinical trials}

\begin{aug}
\author[A]{\fnms{Minge} \snm{Xie}\corref{}\thanksref{t1}\ead[label=e1]{mxie@stat.rutgers.edu}},
\author[A]{\fnms{Regina Y.} \snm{Liu}\thanksref{t2}\ead[label=e2]{rliu@stat.rutgers.edu}},
\author[B]{\fnms{C. V.} \snm{Damaraju}\ead[label=e3]{CDamara1@its.jnj.com}}\break
\and
\author[B]{\fnms{William~H.}~\snm{Olson}\ead[label=e4]{WOlson@its.jnj.com}}

\thankstext{t1}{Supported in part by NSF Grants DMS-11-07012,
DMS-09-15139, SES-0851521 and NSA-H98230-08-1-0104.}
\thankstext{t2}{Supported in part by NSF Grants DMS-10-07683,
DMS-07-07053 and NSA-H98230-11-1-0157.}
\runauthor{Xie, Liu, Damaraju and Olson}
\affiliation{Rutgers, The State University of New Jersey, Rutgers, The State
University of New Jersey,
Ortho-McNeil Janssen Scientific Affairs LLC and Ortho-McNeil Janssen
Scientific Affairs LLC}

\address[A]{M. Xie\\
R. Y. Liu\\
Department of Statistics and Biostatistics\\
501 Hill Center, Busch Campus\\
Rutgers, The State University of New Jersey \\
110 Frelinghuysen Road\\
Piscataway, New Jersey 08854-8019\\
USA\\
\printead{e1}\\
\phantom{E-mail: }\printead*{e2}}
\address[B]{C. V. Damaraju\\
W. H. Olson\\
Quantitative Methodology \\
Ortho-McNeil Janssen\\
\quad Scientific Affairs LLC\\
Raritan, New Jersey 08869\\
USA\\
\printead{e3}\\
\phantom{E-mail: }\printead*{e4}}
\end{aug}

% HISTORY:
\received{\smonth{6} \syear{2011}}
\revised{\smonth{7} \syear{2012}}

% ABSTRACT
%
\begin{abstract}
External information, such as prior information or expert opinions, can
play an important role in the design, analysis and interpretation of
clinical trials.
However, little attention has been devoted thus far to incorporating
external information in clinical trials with binary outcomes, perhaps
due to the perception that binary outcomes can be treated
as normally-distributed outcomes by using normal approximations. In
this paper we show that these two types of clinical trials could behave
differently, and that special care is needed for the analysis of
clinical trials with binary outcomes.
In particular, we first examine a simple but commonly used univariate
Bayesian approach and observe a technical flaw. We then study the full
Bayesian approach using different beta priors and a new frequentist
approach based on the notion of \textit{confidence distribution} (CD).
These approaches are illustrated and compared using data from clinical
studies and simulations.
The full Bayesian approach is theoretically sound,
but surprisingly, under skewed prior distributions,
the estimate derived from the marginal posterior distribution may not
fall between those from the marginal prior and the likelihood of
clinical trial data.
This counterintuitive phenomenon, which we call the ``discrepant
posterior phenomenon,'' does not occur in the CD approach.
The CD approach is also computationally simpler and can be applied
directly to any prior distribution, symmetric or skewed.
\end{abstract}

% KEYWORDS
%
\begin{keyword}
\kwd{Bayesian method}
\kwd{combining information}
\kwd{confidence distribution}
\kwd{expert opinion}
\kwd{requentist method}
\kwd{discrepant posterior}
\kwd{prior distribution}
\end{keyword}

\end{frontmatter}
\newpage
%s1 #&#
%s1 ###
\section{Introduction} \label{sec1}

In pharmaceutical fields as well as many others, there is great
interest in conducting randomized trials with designs that can enable
combining external information, such as prior information or expert
opinions, with trial data to enhance the interpretation of the
findings. In early landmark works \citet{Spi94} and Parmar, Spiegelhalter and Freedman
(\citeyear{Parm}) provided an interesting illustration of
integrating expert opinions with data from cancer trials using a
Bayesian framework. Parmar, Spiegelhalter and Freedman (\citeyear{Parm}) noted that the added
flexibility to such trials to stop for efficacy, futility or safety can
greatly increase the efficiency of clinical research. Designs for
incorporating external information
are also useful in drug development when a pilot study, also known as a
\textit{hypothesis generating study}, is conducted with a sample size that
may be inadequate for detecting clinically meaningful treatment
effects. In this case, relevant information including trial results and
expert opinions can be used to help decision makers with whether to
proceed with a larger confirmatory study and, if so, how to design it.

Although the applications have drawn increasing interest in recent
years, little attention has been devoted to the special yet commonly
seen clinical trials with binary outcomes, partly due to an inaccurate
common belief that little is new regarding the trials with binary
outcomes. In this paper, motivated by a case study of a clinical trial
with binary outcomes in a migraine therapy, we develop and compare
statistical methods which can effectively combine information from
clinical trials of binary outcomes with information from surveys of
expert opinions. The results show that clinical trials with binary
outcomes can behave quite differently.
Thus, special care is warranted for such trials.

The Bayesian paradigm has played a dominant role in combining expert
opinions with clinical trial data. Almost all methods currently used
are Bayesian; see, for example, \citet{Berry}, Spiegelhalter, Abrams and Myles
(\citeyear{spi04}), \citet{Carlin} and \citet{Wij}.
In the Bayesian paradigm, as illustrated in Spiegelhalter, Freedman and Parmar
(\citeyear{Spi94}), a prior distribution is first formed to express the initial
beliefs concerning the parameter of interest based on either some
objective evidence or some subjective judgment or a combination of the
two. Subsequently, clinical trial evidence is
summarized by a likelihood function, and a posterior distribution is
then formed by combining the prior
distribution with the likelihood function. Although this general
Bayesian paradigm also applies to the special case of clinical trials
with binary outcomes, the simple univariate Bayesian approach developed
in Spiegelhalter, Freedman and Parmar (\citeyear{Spi94}) for clinical trials with normally
distributed outcomes cannot be applied directly to clinical trials
with binary outcomes. The latter point is contrary to the common belief
which we elaborate below.\looseness=1

\textit{Univariate Bayesian approach}.

Consider a clinical trial of binary outcomes with a treatment group and
a control group. Denote by $X_{1i}$ $\sim$ $\operatorname{Bernoulli} (p_1), \mbox{ for }  i = 1,\ldots,n_1$,
the\vadjust{\goodbreak} responses from the treatment group
and by $X_{0i} \sim\operatorname{Bernoulli} (p_0), \mbox{ for }  i = 1,\ldots
,\break n_0$, the responses from the control group. Assume that the parameter
of interest is the difference of the success rates between the two
treatments, $\delta\equiv p_1 - p_0$, and its prior distribution $\pi
(\delta)$ is formed based on some objective evidence and/or some
subjective judgment. Let $\hat\delta= {\bar X}_1 - {\bar X}_0$ and
${\hat C}_{d}^2 = {{{\bar X}_1(1- {\bar X}_1)}/{n_1}} + {{{\bar X}_0(1-
{\bar X}_0)}/{n_0}}$, where ${\bar X}_1 = \sum_{i=1}^{n_1} X_{1i}/n_1$
and ${\bar
X}_0 = \sum_{i=1}^{n_0} X_{0i}/n_0$. Note that ${\hat C}_{d}^2$ is an
estimator of $C_d^2 = \operatorname{var} ({\hat\delta}) = {{p_1(1-p_1)}/ {n_1}}
+ {{p_0(1-p_0)}/ {n_0}}.$ A popular univariate Bayesian approach, as
seen in Spiegelhalter, Freedman and Parmar [(\citeyear{Spi94}), pages 360--361], would then treat
%
%e5 #&#
%
%e1.1 ###
\begin{equation}
\label{eqlikeuni} \hat\ell( \delta| \hat\delta) = \exp \bigl\{ -
\tfrac 12\log\bigl(2 \pi\hat C_{d}^2\bigr) - \tfrac1{2} (
\delta- \hat\delta)^2/\hat C_{d}^2 \bigr\}
\end{equation}
as the ``likelihood function'' of $\delta$, and proceed to compute the
posterior distribution of $\delta$, $\hat\pi(\delta| \mathrm{data} )$,
according to
\[
\hat\pi(\delta| \mathrm{data} ) \propto\pi(\delta) \hat\ell(\delta| \hat \delta).
\]

When the prior $ \pi(\cdot) $ is modeled as a normal distribution, the
approach involves an explicit posterior and is straightforward.
Although this univariate Bayesian approach
has been used in practice, it has in fact a theoretical flaw. Strictly
speaking, (\ref{eqlikeuni}) is not a likelihood function of $\delta$,
even in the context of estimated likelihood [see, e.g., \citet
{Boos}]. In particular, in the clinical trial motioned above, a
conditional density function $f(\mathrm{data} |\delta)$ solely depending on a
single parameter $\delta= p_1 - p_0$ does not exist, and, thus, it is
not possible to find a ``marginal likelihood'' of $\delta$. Therefore,
$\pi(\delta| \mathrm{data}) \propto\pi(\delta) f(\mathrm{data}|\delta)$ is not well
defined and any univariate Bayesian approach focusing directly on
$\delta$ is not supported by the Bayesian theory.

The point above is alluded to in the argument made by \citet
{Efr0} and
\citet{Wasser} that a Bayesian approach is not good for
``division of
labor'' in the sense that ``statistical problems need to be solved as
one coherent whole in a Bayesian approach,'' including ``assigning
priors and conducting analyses with nuisance parameters.'' This
observation suggests that a sound Bayesian solution in
the current context is a full Bayesian model that can jointly model
$p_0$ and $p_1$ or their reparametrizations.
Joseph, du Berger and Belisle (\citeyear{Joseph}) presented such a full Bayesian
approach using (mostly) a set of independent beta priors for $p_0$ and~$p_1$. However, the paper focused mainly on the utility of the approach
in sample size determination rather than on its general performance in
the context of clinical trials with binary outcomes. In the present
paper, in addition to the independent beta priors, we broaden the scope
of the full Bayesian approach to include three more flexible priors,
namely, independent hierarchical beta priors, dependent bivariate beta
(BIBETA) priors [\citet{Olk}] and dependent hierarchical
bivariate beta priors.

We also develop a Markov Chain Monte Carlo (MCMC) algorithm for
implementing these full Bayesian approaches, since most resulting
posteriors do not assume explicit forms. The full Bayesian approaches
are theoretically sound, and intuitively would have been expected to
provide a systematic solution to the problems in our case study.
However, a close examination of the situation with skewed priors
reveals a surprising phenomenon in which the estimate derived from the
posterior distribution may not be between those from the prior
distribution and the likelihood function of the observed data (details
in Section~\ref{sec42}). We shall refer to this phenomenon as the ``discrepant
posterior phenomenon.'' To the best of our knowledge, this discrepant
posterior phenomenon has not been reported elsewhere.
This observation indicates that clinical trials of binary outcomes can
behave differently from
the normal clinical trails studied in Spiegelhalter, Freedman and Parmar (\citeyear{Spi94}).
It also shows an inherent difficulty in the modeling of trials with
binary outcomes, especially if $p_0$ and $p_1$ are potentially correlated.
This discrepant posterior phenomenon manifests itself in settings
beyond binary outcomes, and it has far reaching implications in
Bayesian applications in general, as we discuss in Section~\ref{sec5}.

In addition to studying the full Bayesian approach, we also propose a
new frequentist approach for combining external information with
clinical trial data. \citet{Efr0} and \citet{Wasser} argued
that a
frequentist approach has ``the edge of division of labor'' over a
Bayesian approach. They illustrated this point by using the example of
population quantile, which can be directly estimated in a frequentist
setting by its corresponding sample quantile without any modeling
effort or involving other (nuisance) parameters. In our context, this
indicates that we can use a univariate frequentist approach to model
directly the parameter of interest~$\delta$, without having to model
jointly the treatment effects $(p_0, p_1)$. On the other hand, it is
clear that a standard frequentist approach is not equipped to deal with
external information such as expert opinions, which are not actual
observed data from the clinical trials. To overcome this difficulty, we
take advantage of the \textit{confidence distribution} (CD), which uses a
sample-dependent distribution function to estimate a parameter of
interest [see, e.g., \citet{Sch1} and \citet{Sin05}]. In
particular, we use a CD to summarize external
information or expert opinions, and then combine it with the estimates
from the clinical trial. This alternative scheme can be viewed as a
compromise between the Bayesian and frequentist paradigms. It is a
frequentist approach, since the parameter is treated as a fixed value
and not a random entity. It nonetheless also has a Bayesian flavor,
since the prior expert opinions represent only the relative experience
or prior knowledge of the experts but not any actual observed data. The
CD approach is easy to implement and can be a useful data analysis tool
for the type of studies considered in the present paper.

The main emphasis of the paper is on the study and comparison of the
methods for incorporating expert opinions with clinical trial data in
the binary outcome setting, and not the methods for pooling together
individual expert opinions. The latter have been discussed extensively
by \citet{Gen}. The goal of this research is to raise
awareness of the complexity of the practice of incorporating external
information. Although it draws attention to a difference between
Bayesian and non-Bayesian approaches in practice, it is not meant to
either promote or criticize any of the Bayesian or frequentist approaches.

The rest of this section describes a pilot clinical study in a migraine
therapy by \textit{Johnson and Johnson, Inc.} In Section~\ref{sec2} we
develop full
Bayesian approaches with four different priors and implement the
approaches through an MCMC algorithm. In Section~\ref{sec3} we present the
alternative approach of frequentist Bayes compromise using CDs. In
Section~\ref{sec4} we illustrate the approaches discussed in Sections~\ref{sec2} and~\ref{sec3}
using the data presented in Section~\ref{sec11}. We also conduct a simulation
study to compare the performance of these approaches in situations
where the prior distributions are skewed. Finally, we provide in
Section~\ref{sec5} some concluding remarks and discussions.

%s1.1 #&#
%s1.1 ###
\subsection{Application: The pilot study on migraine therapy,
background and data}\label{sec11}

Our data are collected from a recent clinical study on patients with
migraine headaches.\setcounter{footnote}{2}\footnote{Clinical trial NCT00210496 by
Janssen-Ortho LLC (Johnson \& Johnson, Inc.) Web link:
\url{http://clinicaltrials.gov/ct2/results?term=NCT00210496}.}
The objective was to determine the potential impact of a preventive
migraine therapy, topiramate, on the therapeutic efficacy of the acute
migraine therapy, almotriptan.

The study consisted of a 6-week open-label phase followed by a
randomized double-blind phase that lasted 20 weeks. Patients received
topiramate during the open-label run-in period that enabled the
selection for randomization of patients who could tolerate a dosing
regimen of 100~mg$/$day and who met the eligibility criteria based on
migraine rate. Those found eligible were randomly assigned to receive
topiramate (Treatment A) or placebo (Treatment B$/$Control). Throughout
the study, almotriptan 12.5~mg was used as an acute treatment for
symptomatic relief of migraine headaches.
The patients recorded assessments of migraine activity, associated
symptoms and other relevant details into an electronic daily diary
(Personal Digital Assistant [PDA]).
The numbers of patients in the treatment and the control groups are
$n_1 = 59$ and $n_0 = 68$, respectively.
The slight difference in the group size reflects the dropout of a
handful of patients during the double-blind phase. The most common
reason for these dropouts was subject choice/withdrawal of consent. Few
patients discontinued treatment due to limiting adverse event during
the double-blind phase.

%t1 ###
\begin{table}
\tabcolsep=0pt
\caption{Opinion survey of 11 experts on
the treatment improvement $\delta$}\label{tab1}
\begin{tabular*}{\textwidth}{@{\extracolsep{\fill}}lccd{1.2}d{1.2}d{2.2}d{2.2}d{2.2}d{2.2}d{2.2}d{2.2}d{2.2}d{1.2}@{}}
\hline
& \multicolumn{12}{c}{\textbf{Expert opinions for achievement of pain
relief at 2 hours (PR2)}}\\[-4pt]
& \multicolumn{12}{c@{}}{\hrulefill}\\
&\multicolumn{6}{c}{\textbf{Worse (\%)}} & \multicolumn
{6}{c}{\textbf{Better (\%)}}\\[-4pt]
&\multicolumn{6}{c}{\hrulefill} & \multicolumn{6}{c}{\hrulefill}\\
\textbf{Investigator} & \multicolumn{1}{c}{$\bolds{20{+}}$} &
\multicolumn{1}{c}{$\bolds{20{\sim}17}$} & \multicolumn{1}{c}{$\bolds
{16{\sim}13}$} &
\multicolumn{1}{c}{$\bolds{12{\sim}9}$} & \multicolumn{1}{c}{$\bolds
{8{\sim}5}$} &
\multicolumn{1}{c}{$\bolds{4{\sim}0}$} & \multicolumn{1}{c}{$\bolds
{0{\sim}4}$} & \multicolumn{1}{c}{$\bolds{5{\sim}8}$} &
\multicolumn{1}{c}{$\bolds{9{\sim}12}$} & \multicolumn{1}{c}{$\bolds
{13{\sim}16}$} & \multicolumn{1}{c}{$\bolds{17{\sim}20}$} &
\multicolumn{1}{c}{$\bolds{20{+}}$}\\
\hline
\phantom{0}1 & & & 5 & 5 & 10 & 30 & 30 & 15 & 5 & & & \\
\phantom{0}2 & & & & 3 & 7 & 20 & 25 & 20 & 15 & 7 & 3 & \\
\phantom{0}3 & & & & & 10 & 15 & 20 & 20 & 20 & 10 & 5 & \\
\phantom{0}4 & & & 2 & 3 & 5 & 5 & 50 & 20 & 10 & 5 & & \\
\phantom{0}5 & & & & 5 & 10 & 15 & 15 & 30 & 20 & 5 & & \\
\phantom{0}6 & & & & & & 10 & 20 & 30 & 30 & 10 & & \\
\phantom{0}7 & & & & & 5 & 20 & 50 & 20 & 5 & & & \\
\phantom{0}8 & & & & & & 5 & 50 & 40 & 5 & & & \\
\phantom{0}9 & & & & & 5 & 5 & 20 & 30 & 20 & 10 & 10 & \\
10 & & & & & 5 & 15 & 15 & 20 & 15 & 15 & 10 & 5\\
11 & & & & & & 5 & 10 & 30 & 25 & 15 & 10 & 5\\[3pt]
Group mean & & & 0.64 & 1.45 & 5.18 & 13.18 & 27.73 & 25.00 & 15.46 &
7.00 & 3.45 & 0.91\\
\hline
\end{tabular*}
\end{table}

\begin{figure}

\includegraphics{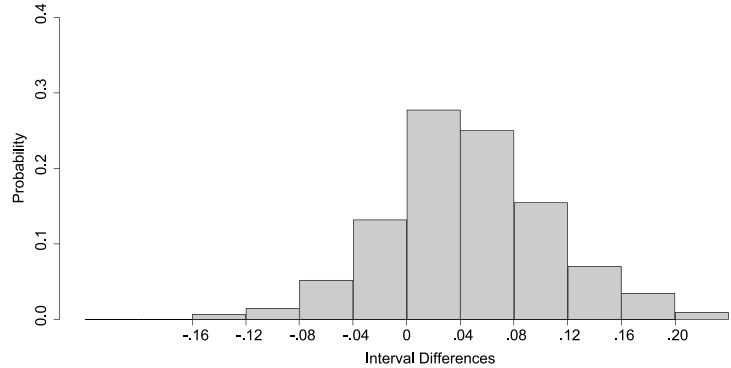}

\caption{Distribution of experts' beliefs (group mean) on the
difference in success rates between
topiramate and control in achieving pain relief at 2 hours in migraine
headaches.}
\label{fig1}
\end{figure}

The trial objectives and study design were presented at an investigator
meeting prior to the start of the study. At the meeting, following the
design of Parmar, Spiegelhalter and Freedman (\citeyear{Parm}), the study sponsor sought the individual
opinions of each investigator expert regarding the expected improvement
of Treatment~A over Treatment~B for a series of clinical outcomes. For
illustration we focus on a specific outcome, pain relief at two hours
(PR2) after dosing with almotriptan, one of 16 outcomes investigated in
the trial.
During the meeting, each expert was asked to use the 12 intervals shown
in Table~\ref{tab1} to assign a ``weight of belief,'' based on his/her
experience, in the difference in the percentage of patients expected to
achieve PR2 in the two treatment groups.
In other words, each expert was given 100 ``virtual patients'' to be
assigned to one of the 12 possible intervals of difference between the
two treatments (from $-20\%$ to $20\%$) in Table~\ref{tab1}.
Table~\ref{tab1} shows the belief distributions for each of the 11
experts and
the group mean. The histogram in Figure~\ref{fig1} shows the group
means of the
11 experts' beliefs of the improvement of Treatment A over Treatment B.

The histogram in Figure~\ref{fig1}, derived from the arithmetic means
in the
last row of Table~\ref{tab1}, is to be used as a (marginal) prior in our
Bayesian analysis for the improvement of Treatment A over Treatment B.
This practice effectively assumes that the heterogeneity of expert
opinions could be averaged out by arithmetic means [cf. \citet
{Gen}]. A similar assumption is also used in Spiegelhalter, Freedman and Parmar (\citeyear{Spi94})
and the development of the frequentist approach in Section
\ref{sec3}. Further discussion on heterogeneity among experts can be
found in
Section~\ref{sec5}.

The goal of our project is to incorporate the information in Figure
\ref{fig1},
solicited from experts, with the data from the pilot clinical trial,
and make inference about the improvement of the treatment effect.
Findings from the inference are intended for generating hypotheses to
be tested in future studies.

%f1 #&#
%
%f1 ###

%%%%%%%%%%%%%%%%%%%%%%%%%%%%%%%%%%%%%%%%%%%%%%%%%%%%%%%%%%%%%%%%%%%%%%%%%%%%%%%%%%%%%%%%%%%%%
%s2 #&#
%s2 ###
\section{A full Bayesian solution: Methodology, theory and algorithm}
\label{sec2}

%s2.1 #&#
%s2.1 ###
\subsection{Summarize external/prior information using an informative
prior distribution}\label{sec21}

Beta distributions are often conventional choices for modeling the
prior of the Bernoulli parameter $p_0$ or $p_1$. They are sufficiently
flexible for capturing distributions of different shapes. In
particular, we consider four forms of joint beta distributions for the
prior of $(p_0, p_1)$.

\begin{itemize}
\item\textit{Independent Beta prior}.
Joseph, du Berger and Belisle (\citeyear{Joseph}) used independent beta priors to summarize
``pre-experimental information'' about ``two independent binomial
parameters'' $p_0$ and $p_1$ as follows: $\pi(p_0, p_1) = \pi(p_0)
\pi
(p_1)$ with $\pi(p_0) \sim\operatorname{BETA}(q_0, r_0)$ and $\pi(p_1) \sim
\operatorname{BETA}(q_1, r_1)$. Here, $(q_0, r_0, q_1, r_1)$ are unknown prior
parameters (hyperparameters) which can be estimated using the method of
moments, following \citet{Lee} and Joseph, du Berger and Belisle (\citeyear{Joseph}). Specifically,
for our clinical study, the average treatment effect and its standard
deviation, $\mu_{d}$ and $\sigma_{d}$, can be obtained (estimated) from
the mean and standard deviation of the histogram of Figure~\ref{fig1}.
Based on
previous clinical trials [cf. \citet{SHNeto3} and \citet{bra}],
the average effectiveness $\mu_0$ of Treatment B and its standard
deviation $\sigma_0$ can also be obtained. We can estimate the prior
parameters $(q_0, r_0, q_1, r_1)$ by solving
the equations $\mu_0 \stackrel{\mathrm{set}}{=} \mathrm{E}(p_0) =
{q_0/{(q_0 + r_0)}}$, $\mu_0 + \mu_d \stackrel{\mathrm{set}}{=} \mathrm{E} (p_1) = {q_1/{(q_1 + r_1)}}$,
$\sigma_0^2 \stackrel{\mathrm{set}}{=} \operatorname{var}(p_0)= {q_0 r_0/\{(q_0 + r_0)^2(q_0 + r_0 + 1)\}}$ and
$\sigma_d^2 - \sigma_0^2 \stackrel{\mathrm{set}}{=} \operatorname{var} (p_1) =
q_1 r_1/\{(q_1 + r_1)^2(q_1 + $
$ r_1 + 1)\}$.

\item \textit{Independent hierarchical Beta prior.}
Gelman et al. [(\citeyear{Gel}), Chapter 5] suggested that hierarchical priors are
more flexible and can avoid ``problems of over-fitting'' in Bayesian models.
We modify their approach to reflect the informative prior in our
problem setting
with two sets of independent Bernoulli experiments.
Specifically, we still model the prior of $(p_0, p_1)$ independently
with $\pi(p_0, p_1) = \pi(p_0) \pi(p_1)$, but each $\pi(p_i)$, for
$i =
0,1$, assumes two levels of hierarchies as follows:
\begin{eqnarray*}
p_i| q_i, r_i &\sim&
\operatorname{BETA}(q_i, r_i),
\\
\xi_i &= &\frac{q_i}{q_i + r_i} \sim\operatorname{BETA}(\alpha_i,
\beta_i) \quad\mbox{and}\quad \eta_i = q_i +
r_i \sim\operatorname{GAMMA}(\alpha_i + \beta_i).
\end{eqnarray*}
Here, $\xi_i$ is the mean and $\eta_i$ is the ``sample size'' of $\operatorname
{BETA}(q_i, r_i)$, following \citet{Gel}, and $\operatorname
{GAMMA}(t)$ refers to the standard gamma distribution whose shape
parameter is $t$ and scale parameter is 1.
Again, we use the method of moments to estimate the unknown parameters
in this (hyper) prior distribution. In this case, the first two
marginal moments of $p_i$ are
$\mathrm{E}(p_i)= \alpha_i/(\alpha_i + \beta_i)$ and $\operatorname
{var}(p_i) =
\alpha_i/\{(\alpha_i + \beta_i)(\alpha_0 + \beta_0 + 1)\}[\alpha_i +
\beta_i + \int\{(x+1)^{-1} x^{\alpha_i + \beta_i} e^{-x}/\Gamma
(\alpha_i + \beta_i)\} \,dx]$, for $i = 0, 1$. The prior parameters $(\alpha_0,
\beta_0, \alpha_1, \beta_1)$ are obtained by solving the equations
$\mu_0 \stackrel{\mathrm{set}}{=} \mathrm{E}(p_0)$, $\mu_0 + \mu_d
\stackrel{\mathrm{set}}{=} \mathrm{E} (p_1)$, $\sigma_0^2 \stackrel{\mathrm{set}}{=}
\operatorname{var}(p_0)$ and
$\sigma_d^2 - \sigma_0^2 \stackrel{\mathrm{set}}{=} \operatorname{var}(p_1)$.

\item\textit{Dependent bivariate Beta (BIBETA) prior}.
Although most of the analysis in Joseph, du Berger and Belisle (\citeyear{Joseph}) was based on the
independent beta prior, a dependent beta prior, with $\pi(p_0)$ and
$\pi
(p_1 | p_0) = \pi(p_0, p_1)/\pi(p_0)$ both being beta distributions,
was considered in a numerical example there. Joseph, du Berger and Belisle (\citeyear{Joseph})
commented that ``it is often desirable to allow dependence between
$p_0$ and $p_1$.'' This point is particularly relevant to our case
study, since almotriptan is used in both groups. However, the
constraint $\mathrm{E}(p_1) = \mathrm{E}(p_0)$, required in the formulation
in Joseph, du Berger and Belisle (\citeyear{Joseph}), does not fit our case. Instead, we use a more
flexible bivariate beta distribution (BIBETA), introduced in
\citet{Olk}, to model $(p_0, p_1)$ in our prior function. This BIBETA
distribution ensures that the marginal prior distributions of $p_0$ and
$p_1$ are both beta distributions. More importantly,
it also allows modeling the correlation between $p_0$ and $p_1$ in the
range [0,1]. This BIBETA distribution, with parameters $q_0$, $q_1$ and
$r$, has a nice latent structure, that is, $p_0 = {U/(U+W)}$ and $p_1 =
{V /(V+W)}$,
where $U, V$ and $W$ are standard gamma random variables with
respective shape parameters $q_0, q_1$ and $r$.
It follows that the joint density (prior distribution) of $p_0$ and
$p_1$ is
\[
\pi(p_0, p_1) \propto \frac{p_0^{q_0 -1} p_1^{q_1 -1} (1 - p_0)^{q_1 +r-1} (1 - p_1)^{q_0 + r-1}} {
(1 - p_0 p_1)^{q_0 + q_1 + r}}.
\]
We obtain the prior parameter values $(q_0, q_1, r)$ by the method of
moments, solving equations
$\mu_0 \stackrel{\mathrm{set}}{=}$ $\mathrm{E}(p_0)= {q_0/{(q_0 +
r)}}$, $
\mu_d \stackrel{\mathrm{set}}{=}$ $\mathrm{E}(p_1 - p_0) = {q_1/{(q_1 +
r)}} - {q_0/{(q_0 + r)}}$ and $
\sigma_d^2 + \mu_d^2 \stackrel{\mathrm{set}}{=}$ $\mathrm{E}(p_1 - p_0)^2
= \{q_0(q_0+1)\}/\{(q_0+r)$ $(q_0+r+1)\}
+ {\{q_1(q_1+1)\}/\{(q_1+r)(q_1+r+1)\}}
- 2_3F_2 (q_0+1, q_1+1, q_0+q_1+r; q_0+q_1+r+1, q_0+q_1+r+1;
1).$
Here $_3F_2(\cdot)$ denotes a hypergeometric function, which can be
calculated using the software \textit{Mathematica}, as mentioned in
\citet{Olk}.

\item\textit{Dependent hierarchical BIBETA prior.}
We also consider a hierarchical BIBETA prior in which we assign
hyperprior distributions on the parameters of the BIBETA distribution:
\begin{eqnarray*}
 (p_0, p_1)| q_0, q_1, r
&\sim&\operatorname{BIBETA}(q_0, q_1, r),
\\
q_0 &\sim&\operatorname{GAMMA}(\alpha_0), q_1
\sim \operatorname{GAMMA}(\alpha_1)\quad \mbox{and}\\
 r& \sim&\operatorname{GAMMA}(\beta).
\end{eqnarray*}
The second level of the hyperprior model implies that
$\xi_i = {q_i/ {(q_i + r)}} \sim\operatorname{BETA}(\alpha_i,$ $ \beta) $ and
$\eta_i = q_i + r \sim\operatorname{GAMMA}(\alpha_i + \beta)$, for $i$ = 0,1,
which matches the conventional parameterization of hierarchical beta priors.
This hierarchical BIBETA prior is more flexible than the regular BIBETA
distribution. To obtain the prior parameter values $(\alpha_0, \alpha_1, \beta)$, we solve equations
$\mu_0 \stackrel{\mathrm{set}}{=} \mathrm{E}(p_0)$, $
\mu_d \stackrel{\mathrm{set}}{=} \mathrm{E}(p_1 - p_0)$ and $\sigma_d^2
\stackrel{\mathrm{set}}{=} \operatorname{var}(p_1 - p_0).$ Here, the marginal
means of $p_0$ and $p_1$ are simply
$\mathrm{E}(p_0)= \alpha_0/(\alpha_0 + \beta)$ and $\mathrm{E}(p_1)=
\alpha_1/(\alpha_1 + \beta)$. The marginal variance $\operatorname{var}(p_1 - p_0) =
\mathrm{E}\{\operatorname{var}(p_1 - p_0 | q_0, q_1, r)\} + \operatorname{var}\{\mathrm{E}(p_1 - p_0 | q_0, q_1, r)\}$ involves three integrations which can
be obtained by numerical integration.
\end{itemize}

%s2.2 #&#
%s2.2 ###
\subsection{Summarize trial data of binary outcomes as a likelihood
function}\label{sec22}
For the clinical trial with binary outcomes, the likelihood
function of $(p_0, p_1)$ is
%e2.1 ###
\begin{equation}
\qquad{\ell}(p_0, p_1| n_0, {\bar X}_0, n_1, {\bar X}_1) \propto p_0^{
n_0 {\bar X}_0} p_1^{n_1 {\bar X}_1}
(1-p_0)^{n_0(1-{\bar X}_0)}
(1-p_1)^{n_1(1-{\bar X}_1)}.
\label{eqbinomlike}
\end{equation}

%s2.3 #&#
%s2.3 ###
\subsection{Combine prior information and trial data as a posterior
distribution}\label{sec23}

Following the Bayes formula, each of the four prior distributions can
be incorporated with the likelihood function (\ref{eqbinomlike}) to
produce a joint posterior distribution of $(p_0, p_1)$,
%
%e6 #&#
%
%e2.2 ###
\begin{equation}
\label{eqpostjoint}\pi(p_0, p_1|
n_0, {\bar X}_0, n_1, {\bar
X}_1) \propto\pi(p_0, p_1)
\ell(p_0, p_1| n_0, {\bar X}_0,
n_1, {\bar X}_1).
\end{equation}
The marginal posterior distribution for the parameter of interest
$\delta(\equiv p_1- p_0)$ is then
%e2.3 ###
\begin{equation}\label{eq3marg}
\qquad\pi(\delta| n_0, {\bar X}_0, n_1, {\bar X}_1) = \int_{\max(0,
-\delta
)}^{\min(-\delta+1, 1)} f(p_0, p_0 + \delta| n_0, {\bar X}_0, n_1,
{\bar X}_1) \,d p_0,
\end{equation}
from which exact Bayesian inferences for $\delta$ can be drawn.

In the case where the prior is modeled by two independent beta
distributions, the posterior distribution (\ref{eqpostjoint}) is
simply a product of two independent beta distributions, $\operatorname{BETA}(n_0 \bar
X_0 + q_0, n_0 (1 - \bar X_0) + r_0)$ and $\operatorname{BETA}(n_1 \bar X_1 + q_1, n_1
(1 - \bar X_1) + r_1)$. However, in the other three cases both (\ref
{eqpostjoint}) and (\ref{eq3marg}) are not of any known form of
distributions and thus are difficult to manipulate.
To this end, we propose a \textit{Metropolis--Hastings algorithm} to
simulate random samples from the posterior distributions (\ref
{eqpostjoint}) and (\ref{eq3marg}). See Appendix~I [\citet
{Xie}] for the proposed Metropolis--Hastings algorithm.

The resulting marginal posterior density of $\delta$ in (\ref
{eq3marg}) incorporates the prior evidence of expert opinions on the
treatment improvement $\delta$ with the evidence from clinical data.
The full Bayesian approaches are theoretically sound and should provide
a systematic solution to our problem in the Bayesian paradigm. However,
as observed in Section~\ref{sec42}, in the case of a skewed prior, the
approaches may lead to the discrepant posterior phenomenon in that the
posterior distributions of $\delta$ can yield an estimate that is not
between the two estimates derived from the corresponding prior
distribution and the likelihood evidence! Further examination suggests
that this phenomenon is quite general. See Section~\ref{sec42} for details.

%%%%%%%%%%%%%%%%%%%%%%%%%%%%%%%%%%%%%%%%%%%%%%%%%%%%%%%%%%%%%%%%%%%%%%%%%%%%%%%%%%%%%%%%%%%%%
%s3 #&#
%s3 ###
\section{A frequentist Bayes compromise: A CD approach} \label{sec3}

In this section we use the so-called \textit{confidence distribution} (CD)
to develop a new approach for incorporating expert opinions with the
trial data of binary outcomes. This approach follows frequentist
principles and treats parameters as fixed nonrandom values. It provides
an attractive alternative to Bayesian methods. In Section~\ref{sec31} we
provide a definition and a brief review of the CD concept. In Sections
\ref{sec32}--\ref{sec34} we develop the proposed CD approach. This
approach can be
simply outlined as follows: use a CD to summarize the prior information
or expert opinions (Section~\ref{sec32}), use another CD (often from a
likelihood function) to summarize the observed data from the clinical
trial (Section~\ref{sec33}), and then combine these two CDs into one CD
(Section~\ref{sec34}). This combined CD can be used to derive various
inferences. Its role in frequentist inference is similar to that of a
posterior distribution in Bayesian inference.
This development provides yet another example in which a CD can provide
useful statistical inference tools for problems where frequentist
methods with desirable properties were previously unavailable.
\citet{Bickel} gives a similar development for normal clinical trials
using an objective Bayes argument. The review article by \citet
{Xie1} contains further discussion.

%s3.1 #&#
%s3.1 ###
\subsection{A brief review of confidence distribution (CD)}\label{sec31}

The CD concept is a simple one. For practical purposes, a CD is simply
a distribution estimator for a parameter of interest. More
specifically, instead of the usual point estimators or interval
estimators (i.e., confidence intervals), CD uses a distribution
function to estimate the parameter.
The development of the CD has a long history; see, for example,
\citet{Fis1}, \citet{Ney} and \citet{Leh}.
But its associated inference schemes and applications have not received
much attention until recently; see, for example, \citet{Efr3}, Schweder
and Hjort (\citeyear{Sch1}, \citeyear{SH2}, \citeyear{SH3}),
Singh, Xie and Strawderman (\citeyear{Sin01}, \citeyear{Sin05}, \citeyear{Sin07}),
\citet{lawless},
\citet{Tia}, \citet{XieSings} and \citet{Sin10}.
Although the CD approach is closely related to Fisher's fiducial
approach, as seen in the classical literature, the new CD developments
are purely frequentist tools involving no fiducial reasoning. Further
discussion of this point as well as the relations between CD-based
inference and fiducial and Bayesian inferences can be found in the
comprehensive review by \citet{Xie1}.

The following CD definition is formulated in \citet{Sch1}
and Singh, Xie and Strawderman (\citeyear{Sin05}) under the framework of frequentist inference.
Singh, Xie and Strawderman (\citeyear{Sin05}) demonstrated that this new version is consistent
with the classical CD definition, and it is easier to use in practice.
In the definition,
$\theta$ (fixed/nonrandom) is the unknown parameter of interest,
$\Theta$ is its parameter space, ${\bX}_n = (X_1,\ldots, $ $ X_n)^T$
is the sample data set, and $\mathcal{X}$ is the corresponding sample space.

\renewcommand{\thedefinition}{\Alph{definition}}
\begin{definition}\label{defA} A function $H_n(\cdot) =
H_n({\bX}_n,\cdot)$ on $\mathcal{X}\times\Theta\to[0,1]$ is a
confidence distribution (CD) for a parameter $\theta$, if it meets the
following two requirements: (R1) For
each given ${\bX}_n\in\mathcal{X}$, $H_n(\cdot)$ is a continuous
cumulative distribution function; (R2) At the true parameter value
$\theta= \theta_0$, $H_n(\theta_0) \equiv H_n({\bX}_n, \theta_0)$, as
a function of the sample ${\bX}_n$, follows the uniform distribution $U[0,1]$.

The function $H_n(\cdot)$ is an asymptotic confidence
distribution (aCD), if the $U[0,1]$ requirement holds only
asymptotically, and the continuity requirement on $H_n(\cdot)$ is
dropped. Also, when it exists, $h_n(\theta) = H_n'(\theta)$ is called a
CD density or confidence density.
\end{definition}

The CD is a function of both the parameter and the random sample. It is
also a sample-dependent distribution function on the parameter space,
following requirement \textit{R1}. Conceptually, it estimates the
parameter by a distribution function.
As an estimation instrument, it is not much different from a point
estimator or a confidence interval. For example, for point estimation,
any single point (a real value or a statistic) can, in principle, be an
estimate for a parameter, and we often impose additional restrictions
to ensure that the point estimator has certain desired properties, such
as unbiasedness, consistency, etc. The two requirements in Definition~\ref{defA}
play roles similar to those restrictions. Specifically, \textit{R1}
suggests that a sample-dependent distribution function on the parameter
space can potentially be used as an estimate for the parameter. The
$U[0,1]$ requirement in \textit{R2} ensures that the statistical
inferences (e.g., point estimates, confidence intervals, $p$-values)
derived from the CD have desired frequentist properties.

Like a posterior distribution function, a CD contains a wealth of
information for inference. It is a useful device for constructing all
types of frequentist statistical inferences, including point estimates,
confidence intervals and $p$-values. For instance, it is evident from
requirement \textit{R2} that intervals obtained from a confidence
distribution such as $(H_n^{-1}(\alpha_1), H_n^{-1}(1 - \alpha_2))$ can
always maintain the nominal level of $100 (1 - \alpha_1 - \alpha_2)\%$
for coverage of $\theta$. See Section 4 of \citet{Xie1} and
references therein for more details. Also,
the CD concept is rather general.
In fact, recent research has shown that Definition~\ref{defA} encompasses a wide
range of existing examples, including most examples in the classical
development of Fisher's fiducial distributions,
bootstrap distributions, significance functions [$p$-value functions,
\citet{Fra}], standardized
likelihood functions, and certain Bayesian prior and posterior
distributions; see, for example, \citet{Sch1}, Singh, Xie and Strawderman (\citeyear{Sin05}, \citeyear{Sin07}) and \citet{Xie1}.

Two examples of CDs which are relevant to the exposition of this paper
are provided in Appendix~II [\citet{Xie}]. Further, Singh, Xie and Strawderman
(\citeyear{Sin05}) and Xie, Singh and Strawderman (\citeyear{XieSings}) developed a general method for combining
CDs from independent studies, which is utilized in Section~\ref{sec34}.

%s3.2 #&#
%s3.2 ###
\subsection{Summarize external/prior information using a CD}\label{sec32}

A key task in our CD approach in this paper is to construct a CD which
summarizes the treatment improvement $\delta$, using only the
information obtained prior to the clinical trial. In the following few
paragraphs we use a set of modeling arguments to justify that the
distribution underlying the histogram in Figure~\ref{fig1} is a CD for
the prior
information. Some of these arguments are similar to those used in
\citet{Gen} for Bayesian approaches, and our concluded
prior CD matches in form the prior distribution suggested by
Spiegelhalter, Freedman and Parmar (\citeyear{Spi94}). This match of our prior CD with the
commonly used Bayesian prior allows a comparison of the CD approach and
the Bayesian approach on an equal footing. Note that what we show here
is only one of many possible modeling approaches to achieve our
purpose. We will not dwell on this topic since the main goal of the
paper is to study and compare inference approaches of incorporating
expert opinions with clinical trial data.

Example~A.2 in Appendix~II [\citet{Xie}] shows that an
informative prior could be viewed as a CD, provided that a sample space
of the prior knowledge or past experiments\vadjust{\goodbreak} can be defined. In the same
spirit, we assume that the expert opinions are based on past knowledge
or experiments about the improvement $\delta$ (the knowledge could be
from experience or from similar, or informal, or even virtual
experiments, but no actual data are available).
This assumption ensures an informative prior and allows us to have a
prior CD for the improvement $\delta$. In particular, let ${\bY}_0$ be
a statistic (with the sample realization $\mathbf{y}_0$)
that summarizes the
information on $\delta$ gathered from past experience or experiments.
Let $\hat\delta({\bY}_0)$ be an estimator of $\delta$ and also let
$F(t) = P \{\hat\delta({\bY}_0) - \delta\leq t \}$ be the
cumulative distribution function of $(\hat\delta({\bY}_0) - \delta)$.
We assume for simplicity that $F(t)$ does not involve unknown nuisance
parameters or, if it does, that they are replaced by their respective
consistent estimates (in this case the development here holds only
asymptotically).
The prior knowledge then gives rise to the following CD (or asymptotic
CD) for~$\delta$:
%
%e7 #&#
%
%e3.1 ###
\begin{equation}
\label{eqcdprior0} H_0(\delta) = 1 - F
\bigl(\hat\delta(\mathbf{y}_0) - \delta\bigr),
\end{equation}
since the two requirements in Definition~\ref{defA} hold for $H_0(\delta)$.
For illustration, consider the case in which $\{\hat\delta({\bY}_0) -
\delta\}/s_0 \to N(0,1)$, where $s_0^2$ is an estimate of $\operatorname
{var}(\hat\delta({\bY}_0))$. In this case, (\ref{eqcdprior0}) is
$H_0(\cdot) = \Phi(\{\cdot- \hat\delta(\mathbf{y}_0)\}/s_0)$.
Equivalently, $N(\hat\delta(\mathbf{y}_0), s_0^2)$ is a
distribution
estimate of $\delta$.

In practice, the realization $\mathbf{y}_0$ of the prior
trials is
unobserved or only vaguely perceived. We rely on a survey of expert
opinions to recover this prior information and $H_0(\delta)$, as in our
case study in Section~\ref{sec11}. For simplicity, we assume that the~$I$
experts in the survey are randomly selected from a large pool of
experts on the subject matter. We also assume that the experts are
randomly exposed to some pre-existing experiments or knowledge, which
in fact resembles a bootstrapping procedure. Denote by ${\bY}_i^*$ the
summary statistic of the pre-existing knowledge on~$\delta$ which the
$i$th expert is exposed to and upon which his/her opinion is based.
It follows that ${\bY}_i^*$ is a bootstrap copy of ${\bY}_0$. Following
Example 2.4 of Singh, Xie and Strawderman (\citeyear{Sin05}), a CD for $\delta$ from the
bootstrap sample ${\bY}_i^*$ is
%
%e8 #&#
%
%e3.2 ###
\begin{equation}
\label{eqboots} H_{0,i}^*(\delta) = 1 - P\bigl\{\hat\delta\bigl({
\bY}_i^*\bigr) - \hat\delta(\mathbf{y}_0) \le\hat\delta
({\bY}_0) - \delta| {\bY}_0\bigr\}.
\end{equation}
This $H_{0,i}^*(\cdot)$ is usually the same as $H_0(\cdot)$ with
probability 1 under some mild conditions, such as those required for
standard bootstrap theory.

However, the function $H_{0,i}^*(\delta)$ only summarizes the prior
knowledge which the $i$th expert is exposed to. We need to associate it
with his/her ``reported'' opinion in the survey table such as in Table
\ref{tab1}.
Let us define, from the $i$th row of Table~\ref{tab1}, an (empirical)
cumulative
distribution function
\[
H_{0,i}^{**}(\delta) = \sum_{k = 1}^{12}
g_{i,k}^{**} \mathbf{1}_{(\delta
\geq L_k)},
\]
where $100 g_{i,k}^{**}$ is the $k$th number reported in the $i$th row
of Table~\ref{tab1}, $L_k$ is the lower bound of the $k$th interval
and ${\bf
1}_{(\cdot)}$ is the indicator function. This $H_{0,i}^{**}(\delta)$ is
the ``reported'' distribution for the improvement $\delta$ by the
$i$th expert.\looseness=-1\vadjust{\goodbreak}

In the ideal case, if the ``reported'' expert opinion faithfully
recorded the ``true'' expert opinion and the ``true'' expert opinion
truly reflects the ``true'' prior knowledge, $H_{0,i}^*(\delta)$ and
$H_{0,i}^{**}(\delta)$ would be the same. But, there are often
variations in reality.
A detailed discussion on how to model such variations is provided in
Section~\ref{sec5} as a concluding remark. We proceed with the popular
``arithmetic pooling'' approach, which is also articulated in
Spiegelhalter, Freedman and Parmar (\citeyear{Spi94}).
An underlying assumption of arithmetic pooling is that the average of
the ``observed'' expert opinions is an unbiased representation of the
``true'' prior knowledge. In our case, this is equivalent to assuming
the additive error model, $H_{0,i}^{**}(\delta) = H_{0,i}^*(\delta) +
e_i$ such that
$I^{-1} \sum_{i = 1}^I e_i \approx0$ uniformly in $\delta$, where $e_i
= e_i(\delta)$ is defined as the difference between
$H_{0,i}^{**}(\delta
)$ and $H_{0,i}^*(\delta)$ and is viewed as a random error for both the
discrepancies between the ``true'' prior knowledge, the ``true'' expert
opinion and the ``reported'' opinion of the $i$th expert.
Under this error model, where the heterogeneous deviation among experts
are ``averaged out,'' it follows that
%
%e9 #&#
%
%e3.3 ###
\begin{equation}
\label{eqaveraging} H_0(\delta) \approx I^{-1} \sum
_{i = 1}^I H_{0,i}^{**}(
\delta) \approx \sum_{k = 1}^{12} \bar
g_k \mathbf{1}_{(\delta\geq L_k)},
\end{equation}
where $ 100 \bar g_k = 100 I^{-1} \sum_{i = 1}^I g_{i,k}^{**}$
are the group means reported in the last row of Table~\ref{tab1}. From
(\ref
{eqaveraging}), the (standardized) histogram in Figure~\ref{fig1},
that is,
$f_{\mathrm{hist}}(\delta) = \sum_{k = 1}^{12}$ $\{\bar g_k/{(L_{k+1} - L_k)}
\}
$ $\mathbf{1}_{(L_{k+1} > \delta\geq L_k)}$, is clearly a suitable
approximation for the underlying confidence density function
$h_0(\delta
) = H_0'(\delta)$.
Here, the word ``standardized'' refers to scaling the histogram so that its
area is 1. In our calculations in Section~\ref{sec4}, we have used
$L_{13} =
0.24$ as the upper bound of the $12$th interval.

%s3.3 #&#
%s3.3 ###
\subsection{Summarize clinical trial data as a CD}\label{sec33}

The task to summarize the clinical trial results into a CD is
relatively easier. The maximum likelihood estimator of $\delta$ is
${\hat\delta} = {\bar X}_1 -
{\bar X}_0$ with variance $C_d^2 = \operatorname{var} ({\hat\delta}) =
{{p_1(1-p_1)}/ {n_1}} + {{p_0(1-p_0)}/ {n_0}}.$ An estimator of
$C_{d}^2$ is ${\hat C}_d^2 = {{{\bar X}_1(1- {\bar
X}_1)}/{n_1}} +
{{{\bar X}_0(1- {\bar X}_0)}/{n_0}}.$
If both $n_i$'s tend to $\infty$, we have ${{({\hat\delta} -
\delta)}/{ {\hat C}_{d}}} \to N(0,1)$. Therefore, an asymptotic CD for
the parameter $\delta$ from the
clinical trial is
%
%e10 #&#
%
%e3.4 ###
\begin{equation}
\label{eqHtnormal} H_T(\delta) = \Phi \bigl({({
\delta- {\hat\delta}})/ { {\hat C}_{d}}} \bigr).
\end{equation}
In other words, the distribution $N({\hat\delta}, {\hat C}_{d}^2)$
can be used to estimate $\delta$.
An alternative approach is to use the profile likelihood function of
$\delta$. Specifically, let $\ell_\mathrm{prof}(\delta)$ be the log
profile likelihood function of $\delta$, and let $\ell_n^*(\delta) =
\ell_\mathrm{prof}(\delta) - \ell_\mathrm{prof}(\hat\delta)$. Following
Singh, Xie and Strawderman (\citeyear{Sin07}), we can show that
%
%e11 #&#
%
%e3.5 ###
\begin{equation}
\label{eqHtlog}H_T(\delta) = \int
_{-1}^{\delta} h_T(\theta) \,d \theta\qquad
\mbox{with } h_T(\theta) = \frac{e^{\ell_n^*(\theta)}} {\int_{-1}^1 e^{\ell_n^*(\theta)} \,d \theta} \propto e^{\ell_\mathrm{prof}(\delta)}
\end{equation}
is an asymptotic CD for $\delta$. The two $H_T(\cdot)$ above are
asymptotically equivalent when the $n_i$'s tend to $\infty$.

%s3.4 #&#
%s3.4 ###
\subsection{Combine prior information and trial data as a combined
CD}\label{sec34}

We can incorporate the prior CD $H_0(\delta)$ with the CD
$H_T(\delta)$ from the clinical trial using a general CD combination
method developed by
Singh, Xie and Strawderman (\citeyear{Sin05}). Xie, Singh and Strawderman (\citeyear{XieSings}) showed that this general method
and its extension can provide a unifying framework for most information
combination methods used in current practice, including both the
classical approach of combining $p$-values and the modern model-based
(fixed and random effects models) meta-analysis approach.
In our context, we are combining two CDs. Specifically, we let $G_c(t)
= P(g_c(U_1, U_2) \le t)$, where $U_1$ and $U_2$ are independent
U$[0,1]$ random variables, and $g_c(u_1,u_2)$ is a continuous function
from $[0,1]\times[0,1]$ to $\real$ which is monotonic (say,
increasing) in
each coordinate. Then,
%
%e12 #&#
%
%e3.6 ###
\begin{equation}
\label{eqH-comb} H^{(c)}(\delta) = G_c\bigl
\{g_c\bigl(H_0(\delta),H_T(\delta)\bigr)
\bigr\}
\end{equation}
is a combined CD for $\delta$ which contains information from both
expert opinions and the clinical
trial.
One simple choice is
%
%e13 #&#
%
%e3.7 ###
\begin{equation}
\label{eqg-weight} g_c(u_1,u_2) =
w_1 \Phi^{-1}(u_1) + w_2
\Phi^{-1}(u_2),
\end{equation}
with weights $w_1 = 1/{\hat\sigma_d}$ and $w_2 = 1/ \hat C_{d}$, where
$\hat\sigma_d$ is an estimate of the standard deviation of
$H_0(\delta)$
(specifically, $\hat\sigma_d$ is the standard deviation of the
histogram in Figure~\ref{fig1} in our application, and it is also an
estimate of
the standard
deviation of $\hat\delta$ in the normal case).
In this case, $G_c(t) $ can be expressed as $ \Phi (t / \{
(1/\hat\sigma_d^2 +
1/\hat C_{d}^2)^{1/2} \} )$, and thus gives rise to the
following combined CD for $\delta$: $H^{(c)}(\delta) = \Phi
( \{
\Phi^{-1}(H_0(\delta))/\hat\sigma_d+
\Phi^{-1}(H_T(\delta))$ $/\hat C_{d} \} / \{(1/\hat
\sigma_d^2 +
1/\hat C_{d}^2)^{1/2} \} )$.

When both $H_0(\delta)$ and $H_T(\delta)$ are normal (or asymptotically
normal) CDs, the normal combination in (\ref{eqg-weight}) is the most
efficient in terms of preserving Fisher information.
In nonnormal cases, Singh, Xie and Strawderman (\citeyear{Sin05}) studied several choices of the
function $g_c$ and their Bahadur efficiency. But it remains an open
question what choice of $g_c$ is most efficient in preserving Fisher
information in a general nonnormal setting. Although we use the simple
normal combination in~(\ref{eqg-weight}) in this paper mostly for simplicity,
our experience with the numerical studies has shown this combination to
be quite adequate in most applications. In fact, in many nonnormal
cases, it incurs very little loss of efficiency in terms of preserving
Fisher information from both $H_0(\delta)$ and $H_T(\delta)$.

In a Bayesian approach, it is a conventional practice to fit a prior
density curve to the histogram in Figure~\ref{fig1}. Although this
step is not
needed in the proposed CD approach, we may sometimes also fit a density
function to $f_{\mathrm{hist}}(\delta)$. For example, we may fit a normal curve
to the histogram of Figure~\ref{fig1} by matching its first two
moments, say,
mean $\hat\mu_d$ and variance $\hat\sigma_d^2$. In this case, we have
a normal CD from the expert opinions $H_0(\delta)\approx\Phi((\delta-
\hat\mu_d)/\hat\sigma_d)$.
Incorporating it with $H_T(\delta)$ in~(\ref{eqHtnormal}), we have
the combined CD
$H^{(c)}(\delta)= \Phi ((\delta- \tilde\delta)/{\tilde
C_{d}}
)$ or $N({\tilde\delta}, \tilde C_d^2),$
where $\tilde\delta= ({\hat\delta}/{\hat C}_{d}^2+ \hat\mu_{d}/\hat
\sigma_d^2)/
(1/{\hat C}_{d}^2 + 1/\hat\sigma_d^2)$ and
$\tilde C_{d}^2 = (1/{\hat C}_{d}^2 + 1/\hat\sigma_d^2)^{-1}.$
This combined CD turns out to be the same as the posterior distribution
function obtained from the univariate Bayesian approach described in
the \hyperref[sec1]{Introduction} when the normal prior $\pi(\delta) \sim N(\hat\mu_d,
\hat\sigma_d^2)$ is used. Because Bayes's formula requires that we
know $f(\mathrm{data}|\delta)$ in the univariate Bayesian approach and this
conditional density function $f(\mathrm{data}|\delta)$ does not exist in our
clinical setting,
we have argued that the univariate Bayesian approach is not supported
by Bayesian theory.
The CD development, interestingly, provides theoretical support for
using the posterior distribution
$N(\tilde\delta, \tilde C_d^2)$ from a non-Bayesian point of view if
the prior distribution of expert opinions can be approximated by a
normal distribution. In this case, the univariate Bayesian approach can
also produce a result that ``makes sense,'' and practically we can use
either the CD approach or the univariate Bayesian approach. But this
statement is not true in general.

%%%%%%%%%%%%%%%%%%%%%%%%%%%%%%%%%%%%%%%%%%%%%%%%%%%%%%%%%%%%%%%%%%%%%%%%%%%%%%%%%%%%%%%%%%%%%

%f2 #&#
%
%f2 ###
\begin{figure}

\includegraphics{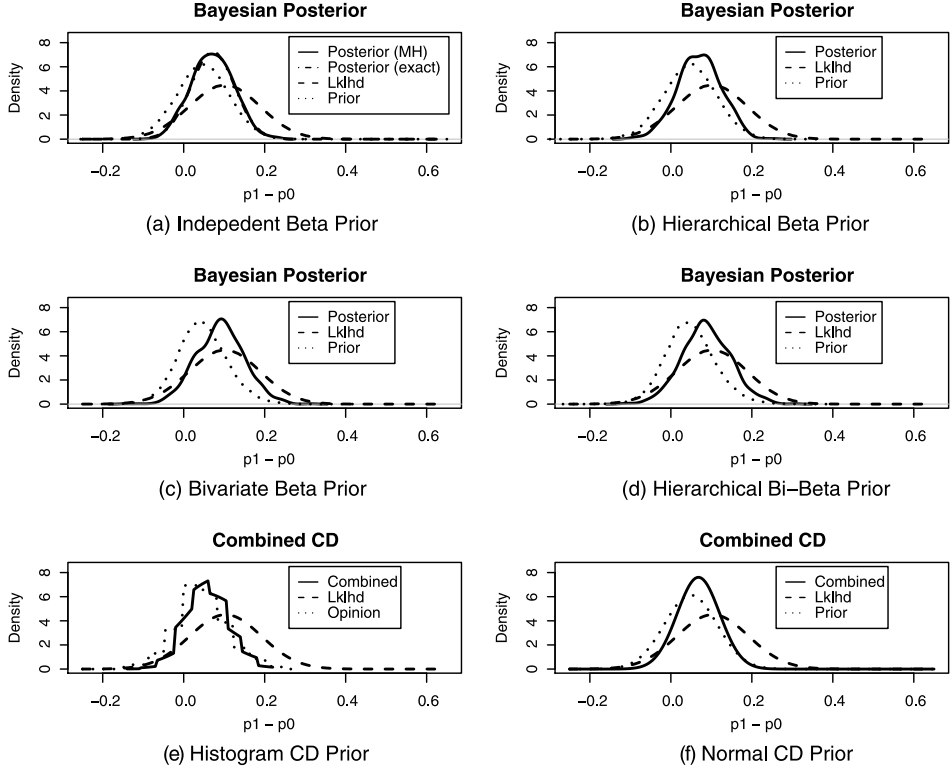}

\caption{Outcomes of data analysis from the migraine pain data.}\label{fig2}\vspace*{-3pt}
\end{figure}

%t2 ###
\begin{table}
\tabcolsep=0pt
\def\arraystretch{0.9}
\caption{Numerical results from incorporating expert opinions
on PR2 (summarized in Figure \protect\ref{fig1}) with clinical data
on PR2: mode,
median, mean,
$I_{80\%}$, $I_{90\%}$ and $I_{95\%}$ of the marginal prior,
(normalized) profile likelihood function and marginal posterior of the
parameter $\delta$. Here, $I_{80\%}$, $I_{90\%}$ and $I_{95\%}$ denote
the interval [$100\alpha$\%-tile, $100(1-\alpha)$\%-tile] for $\alpha=
10\%$, $5\%$ and $2.5\%$, respectively.
Included in the comparison are four full Bayesian approaches and two
approaches based on confidence distributions (CDs)}\label{tab2}
{\fontsize{8.7}{10.7}\selectfont{
\begin{tabular*}{\textwidth}{@{\extracolsep{\fill
}}lccccd{2.8}d{2.8}d{2.8}@{}}
\hline
& & \textbf{Mode} & \textbf{Median} & \textbf{Mean} & \multicolumn
{1}{c}{$\bolds{I_{80\%}}$} & \multicolumn{1}{c}{$\bolds{I_{90\%}}$}
&\multicolumn{1}{c@{}}{$\bolds{I_{95\%}}$} \\
\hline
\multicolumn{8}{c}{Bayesian approaches}\\
Ind Beta prior & Prior & 0.049 & 0.047 & 0.048 & -0.037\ 0.130 &
-0.060\ 0.153 &
-0.080\ 0.173 \\
& Likelihood & 0.104 & 0.104 & 0.103 & -0.007\ 0.214 & -0.043\ 0.251 &
-0.062\ 0.269 \\
& Posterior & 0.069 & 0.070 & 0.071 & 0.000\ 0.138 & -0.015\ 0.156 &
-0.029\ 0.174
\\[3pt]
Hierarchical & Prior & 0.048 & 0.047 & 0.048 & -0.036\ 0.130 & -0.059\
0.154 &
-0.080\ 0.175 \\
\quad Beta prior & Likelihood & 0.104 & 0.104 & 0.103 & -0.007\ 0.214 &
-0.043\ 0.251
& -0.062\ 0.269 \\
& Posterior & 0.082 & 0.071 & 0.070 & 0.000\ 0.143 & -0.021\ 0.159 &
-0.037\ 0.171
\\[3pt]
Bi-Beta prior & Prior & 0.040 & 0.044 & 0.048 & -0.029\ 0.125 & -0.050\
0.151 &
-0.069\ 0.174 \\
& Likelihood & 0.104 & 0.104 & 0.103 & -0.007\ 0.214 & -0.043\ 0.251 &
-0.062\ 0.269 \\
& Posterior & 0.093 & 0.091 & 0.091 & 0.015\ 0.165 & -0.004\ 0.190 &
-0.025\ 0.209
\\[3pt]
Hierarchical & Prior & 0.043 & 0.045 & 0.048 & -0.028\ 0.127 & -0.049\
0.154 &
-0.068\ 0.178 \\
\quad Bi-Beta prior & Likelihood & 0.104 & 0.104 & 0.103 & -0.007\ 0.214 &
-0.043\ 0.251 & -0.062\ 0.269 \\
& Posterior & 0.082 & 0.086 & 0.087 & 0.013\ 0.162 & -0.01\ \phantom{0}0.189 &
-0.031\ 0.207
\\[6pt]
\multicolumn{8}{c}{CD approaches} \\
CD with & Prior CD & 0.020 & 0.060 & 0.048 & -0.023\ 0.142 & -0.068\
0.145 &
-0.070\ 0.182 \\
\quad histogram prior & Likelihood & 0.104 & 0.104 & 0.103 & -0.007\ 0.214 &
-0.043\ 0.251 & -0.062\ 0.269 \\
& Comb. CD & 0.060 & 0.065 & 0.058 & 0.013\ 0.141 & -0.022\ 0.145 &
-0.025\ 0.182
\\[3pt]
CD with & Prior CD & 0.048 & 0.048 & 0.048 & -0.034\ 0.132 & -0.058\
0.156 &
-0.078\ 0.177 \\
\quad normal prior & Likelihood & 0.104 & 0.104 & 0.103 & -0.007\ 0.214 &
-0.043\ 0.251 & -0.062\ 0.269 \\
& Comb. CD & 0.068 & 0.068 & 0.068 & 0.001\ 0.135 & -0.018\ 0.154 &
-0.035\ 0.171
\\
\hline
\end{tabular*}
}}\vspace*{-3pt}
%
%}
\end{table}
%

%s4 #&#
%s4 ###
\section{Application: Numerical results and comparisons}\label{sec4}

We now provide numerical studies to illustrate and compare the Bayesian
and CD approaches discussed in Sections~\ref{sec2} and~\ref{sec3}. In
Section~\ref{sec41} we
focus on the data from the migraine pain study outlined in Section \ref
{sec11}.
In Section~\ref{sec42} we simulate
a skewed distribution of expert opinions and combine the simulated
prior information with the clinical trial data.

%s4.1 #&#
%s4.1 ###
\subsection{Normal priors: A case study of the migraine pain
data}\label{sec41}

For the outcome PR2, the clinical data
report that 31 out of 68 patients in the control group and 33 out of
59 patients in the treatment group achieved pain
relief at 2 hours. Our goal is to incorporate the expert inputs
reported in Figure~\ref{fig1} with these observed outcomes.

We apply the full Bayesian approach described in Section~\ref{sec2} to analyze
the PR2 data, using each of the four beta priors for $(p_0, p_1)$. The
numerical results are reported in Figures~\ref{fig2}(a)--(d) and also
in the
first 12 rows of Table~\ref{tab2}.
The dotted lines in Figures~\ref{fig2}(a)--(d) indicate the marginal prior
density functions, and the dashed lines indicate the (standardized)
profile likelihood function of $\delta$ based only on the clinical
trial data.
The solid lines in Figures~\ref{fig2}(a)--(d) are the marginal posterior
distributions of $\delta$. They are obtained by using the density
estimation function \textit{density}($\cdot$) in the R software and from 1000
Metropolis--Hasting samples of $\delta^* = p_1^* - p_0^*$. In each case
and for each of the 1000 replications, the Metropolis--Hastings
algorithm is iterated $t = 25\mbox{,}000$ times (burn-in). The acceptance
rates are on average $0.0379$, $0.0381$, $0.0480$ and $0.0485$,
respectively, in (a) to (d). For the independent beta prior, the exact
formula of the posterior distribution is available, and it is plotted
in Figure~\ref{fig2}(a) as the dash-dot broken curve (it is barely
visible in
the plot, since it is almost identical to the solid curve). The close
agreement of
these two curves for the posterior distribution indicates that the MCMC
chain of the Metropolis--Hastings algorithm has generally converged with
$t = 25\mbox{,}000$ in this case.

In applying the CD approach, we use both the raw histogram in Figure~\ref{fig1}
and the $N(\hat\mu_d, \hat\sigma_d^2)$ distribution to approximate
the prior CD of expert opinions.
Figures~\ref{fig2}(e)--(f) and the last six rows of Table~\ref{tab2}
contain numerical results.
The dotted lines in Figures~\ref{fig2}(e)--(f) indicate the prior CDs, the
dashed lines indicate the profile likelihood function of $\delta$ based
only on the clinical trial data, and the solid lines are for the
combined CDs for $\delta$.

In this particular example, all six approaches (four Bayesian and two
CD approaches) seem to yield similar posterior or combined CD
functions, and thus similar statistical inferences, regardless of which
approach is used. Although the six marginal posterior or combined CD
distributions are slightly different from one another, the difference
appears to all fall within the expected estimation error of the density
curves. This result is not surprising, since, although skewed, the
degree of skewness of the histogram in Figure~\ref{fig1} does not
appear to be
great enough to render the normal approximation invalid. In fact, in
this case, such a result is expected to hold if the central limit
theory is in place for the clinical data of binary outcomes. It is
worth noting here that the Bayesian approach implemented through an
MCMC method is more demanding computationally.

%s4.2 #&#
%s4.2 ###
\subsection{Skewed priors: A simulation study}\label{sec42}

The outcome in the previous subsection begs the question of whether
there would be a significant difference among the approaches if the
distribution of expert opinions were unambiguously skewed, so that the
normal approximation is clearly not valid.
Conventional wisdom suggests that full Bayesian approaches based on
beta priors, though computationally more intensive, would have
advantages due to their flexibility in capturing distributions of
various shapes. The CD approaches, allowing skewed priors, may also
work. However, the numerical results reveal a surprising finding in the
full Bayesian approaches.

\begin{figure}

\includegraphics{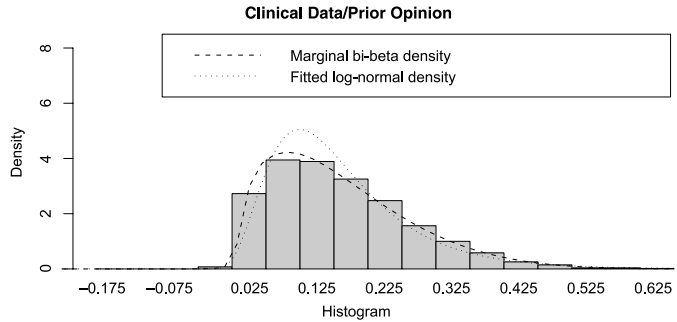}

\caption{Simulated prior distribution of $\delta= p_1 - p_0$ using
$\operatorname{BIBETA}(6, 20, 2)$ distribution.}
\label{fig3}\vspace*{-3pt}
\end{figure}

In this simulation study, we again use the observed clinical data on
PR2, but replace Table~\ref{tab1} and Figure~\ref{fig1} of expert
opinions with their
simulated counterparts, assuming that the underlying prior distribution
function of $(p_0, p_1)$ is a bivariate beta distribution, $\operatorname{BIBETA}(6,
20, 2)$. The marginal means of the $\operatorname{BIBETA}(6, 20, 2)$ distribution are
$\mathrm{E} p_0 = 6/(6 +2)= 0.75$ and $\mathrm{E} p_1 = 20/(20 + 2) \approx
0.90$. Thus, the simulated prior represents a treatment effect
improvement on average about 75\% to 90\%, which are similar to those
of the real trial in Section~\ref{sec41}. Specifically, we simulate
responses of
100 patients for each of the 11 experts from $\operatorname{BIBETA}(6, 20, 2)$, tally
the results in the format of Table~\ref{tab1} (not shown), and then
plot them as
a histogram in Figure~\ref{fig3}.
For a direct visual comparison, Figure~\ref{fig3} includes the curve
of the
$\operatorname{BIBETA}(6, 20, 2)$ density. Also plotted in Figure~\ref{fig3} is, as a common
approach to fitting a skewed distribution, the following fitted
log-normal density $\phi (\{\log(\delta) - \log(\hat\mu_d -
c)\}/\{
1 + \hat\sigma_d^2/(\hat\mu_d - c)^2\}^{1/2} - \frac12 \sqrt{1
+\hat\sigma_d^2/(\hat\mu_d - c)^2} )/\{\delta(1 + \hat\sigma_d^2/(\hat\mu_d - c)^2)^{1/2}\}$. Here, $\hat\mu_d$ and $\hat
\sigma_d$ are the mean and the standard deviation computed from the
histogram, and $c$ is a constant used to capture the shift of the
log-normal distribution from $0$.

%f3 #&#
%
%f3 ###

We apply the same four full Bayesian approaches used in Section \ref
{sec41} to
incorporate
the simulated expert opinions represented in Figure~\ref{fig3} with
the clinical
trial data on PR2. The four sets of prior parameters used in these four
approaches are $(q_0, r_0, q_1, r_1) = (14.66, 4.88, 46.81,4.68)$,
$(\alpha_0, \beta_0, \alpha_1, \beta_1) =\break (30.19, 10.06, 96.43, 9.43)$,
$(q_0, q_1, r) = (6,20,2)$ and $(\alpha_0, \alpha_1, \beta) =
(17.88, $
$59.60, $ $5.96)$, respectively. In the third approach, we directly use
the true set of prior parameters $(q_0, q_1, r) = (6,20,2)$; in the
other three, the prior parameters are obtained by the method of moments
outlined in Section~\ref{sec2}.
In the simulation example, the Metropolis--Hasting algorithm is again
iterated $t = 25,000$ times (burn-in), and it is repeated 1000 times
to obtain 1000 independent Metropolis--Hasting samples of $(p_0^*,
p_1^*)$ in each case. The acceptance rates are on average $0.0019$,
$0.0027$, $0.0057$ and $0.0036$, respectively.\vadjust{\goodbreak}

%f4 #&#
%
%f4 ###
\begin{figure}[b]

\includegraphics{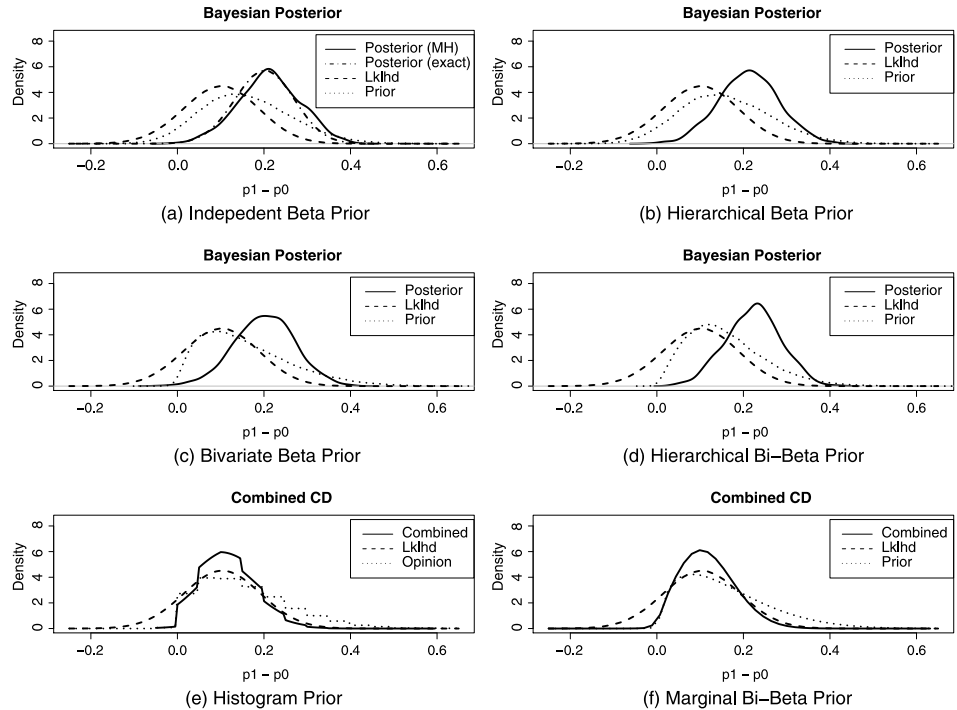}

\caption{Outcomes of data analysis from simulated data with a skewed
prior. }
\label{fig4}
\end{figure}

%t3 ###
\begin{table}
\tabcolsep=0pt
\caption{Numerical results from incorporating the simulated
expert opinions (summarized in Figure \protect\ref{fig3}) with
clinical data on PR2: the
mode, median, mean,
$I_{80\%}$, $I_{90\%}$ and $I_{95\%}$ of the marginal prior,
(normalized) profile likelihood function, and marginal posterior\break of the
parameter $\delta$. Here, $I_{80\%}$, $I_{90\%}$ and $I_{95\%}$ denote
the interval [$100\alpha$\%-tile, $100(1-\alpha)$\%-tile] for $\alpha=
10\%$, $5\%$ and $2.5\%$, respectively.
The prior parameters\break in the four full Bayesian approaches are $(q_0,
r_0, q_1, r_1) = (14.66, 4.88, 46.81,4.68)$, $(\alpha_0, \beta_0,
\alpha_1, \beta_1) = (30.19,10.06, 96.43, 9.43)$, $(q_0, q_1, r) = (6,20,2)$
and\break  $(\alpha_0, \alpha_1, \beta) = (17.88, 59.60, 5.96)$,
respectively}\label{tab3}
{\fontsize{8.7}{10.7}\selectfont{
\begin{tabular*}{\textwidth}{@{\extracolsep{\fill
}}lccccd{2.8}d{2.8}d{2.8}@{}}
\hline
& & \textbf{Mode} & \textbf{Median} & \textbf{Mean} & \multicolumn
{1}{c}{$\bolds{I_{80\%}}$} & \multicolumn{1}{c}{$\bolds{I_{90\%}}$}
&\multicolumn{1}{c@{}}{$\bolds{I_{95\%}}$} \\
\hline
\multicolumn{8}{c}{Bayesian approaches}\\
Independent & Prior & 0.128 & 0.153 & 0.159 & 0.033\ 0.297 & 0.003\
0.340 &
-0.023\ 0.379 \\
\quad Beta prior & Likelihood & 0.104 & 0.104 & 0.103 & -0.007\ 0.214 &
-0.043\ 0.251
& -0.062\ 0.269 \\
& Posterior & 0.211 & 0.212 & 0.212 & 0.120\ 0.306 & 0.089\ 0.330 &
0.066\ 0.346 \\[3pt]
Hierarchical & Prior & 0.145 & 0.152 & 0.159 & 0.031\ 0.295 & -0.001\
0.337 &
-0.027\ 0.375 \\
\quad Beta prior & Likelihood & 0.104 & 0.104 & 0.103 & -0.007\ 0.214 &
-0.043\ 0.251
& -0.062\ 0.269 \\
& Posterior & 0.214 & 0.212 & 0.212 & 0.122\ 0.302 & 0.094\ 0.329 &
0.078\ 0.348 \\[3pt]
Independent & Prior & 0.095 & 0.140 & 0.159 & 0.042\ 0.306 & 0.027\
0.360 &
0.017\ 0.407 \\
\quad Bi-Beta prior & Likelihood & 0.104 & 0.104 & 0.103 & -0.007\ 0.214 &
-0.043\ 0.251 & -0.062\ 0.269 \\
& Posterior & 0.202 & 0.203 & 0.201 & 0.112\ 0.288 & 0.084\ 0.315 &
0.061\ 0.339 \\[3pt]
Hierarchical & Prior & 0.120 & 0.146 & 0.159 & 0.056\ 0.281 & 0.039\
0.326 &
0.028\ 0.366 \\
\quad Bi-Beta prior & Likelihood & 0.104 & 0.104 & 0.103 & -0.007\ 0.214 &
-0.043\ 0.251 & -0.062\ 0.269 \\
& Posterior & 0.232 & 0.225 & 0.222 & 0.138\ 0.305 & 0.116\ 0.329 &
0.101\ 0.340
\\[6pt]
\multicolumn{8}{c}{CD approaches} \\
CD with & Prior CD & 0.075 & 0.125 & 0.159 & 0.025\ 0.275 & -0.025\
0.325 &
-0.025\ 0.375 \\
\quad histogram  & Likelihood & 0.104 & 0.104 & 0.103 & -0.007\ 0.214 &
-0.043\ 0.251 & -0.062\ 0.269 \\
\quad prior & Comb. CD & 0.100 & 0.110 & 0.118 & 0.035\ 0.200 & 0.000\ 0.225 &
-0.005\ 0.250 \\[3pt]
CD with & Prior CD & 0.095 & 0.140 & 0.159 & 0.042\ 0.306 & 0.027\
0.360 &
0.017\ 0.407 \\
\quad marginal  & Likelihood & 0.104 & 0.104 & 0.103 & -0.007\
0.214 &
-0.043\ 0.251 & -0.062\ 0.269 \\
\quad Bi-Beta prior& Comb. CD & 0.099 & 0.099 & 0.119 & 0.026\ 0.191 & 0.007\ 0.209 &
-0.011\ 0.246 \\
\hline
\end{tabular*}
}}
%
%}
\end{table}

We also apply two CD combination approaches to incorporate
the simulated expert opinions in Figure~\ref{fig3} with the clinical
trial data
on PR2. Similar to that in Section~\ref{sec4}, the first CD approach directly
uses the raw histogram in Figure~\ref{fig3}. The second CD approach,
in order to
have a direct comparison with the Bayesian approach using the
underlying prior $\operatorname{BIBETA}(6, 20, 2)$, combines the underlying marginal
density function of $\delta$ with the CD from the clinical trial data.
Of course, in reality we do not know the underlying prior distribution
or the underlying marginal density function of $\delta$. Thus, the
second CD approach has only theoretical value. Without relying on the
underlying CD prior, we also consider the CD approach which combines
the fitted log-normal distribution in Figure~\ref{fig3} with the CD
from the
clinical trial data. However, since the log-normal curve is evidently a
poor fit for the histogram in Figure~\ref{fig3}, the result for this CD
approach, though not too far off, does not seem well justified and is
thus omitted.

Figures~\ref{fig4}(a)--(d) show the results on the improvement $\delta
$ using
the full Bayesian approaches, and Figures~\ref{fig4}(e)--(f) show the results
using the CD combination approaches. Figure~\ref{fig4} adopts the same notation
and symbols used in Figure~\ref{fig2}. Again, for the independent beta prior,
the posterior density from the algorithm closely matches the one using
its exact formula (dashed-dotted line), indicating that the MCMC chain
of the Metropolis--Hasting algorithm has generally converged in this
case. Also, we report in Table~\ref{tab3} the numerical results from
the six
approaches: the mode, median, mean and confidence/credible intervals of
the marginal priors, the profile likelihood function and the
marginal posteriors of $\delta$.

The CD approaches perform exactly as anticipated. However, examining
the modes of the three curves in each of Figures~\ref{fig4}(a)--(d), we notice
that the mode of the marginal posterior distribution (solid curve) lies
to the right of the peaks of both the marginal prior distribution
(dotted curve) and the profile likelihood function (dashed curve). The
numerical results in Table~\ref{tab3} also confirm that the mode,
median and
mean of the marginal posterior distributions of $\delta$ from all four
full Bayesian approaches are much larger than their counterparts from
the corresponding marginal priors and profile likelihood functions.
This discrepant posterior phenomenon is counterintuitive! For instance,
if we use the means as our point estimators, we would report from
Figure~\ref{fig4}(c) that the experts suggest about $15.9\%$
improvement and the
clinical evidence suggests about $10.3\%$ improvement but,
incorporating them together, the overall estimator of the treatment
effect is $20.1\%$, which is bigger than either that reported by the
experts or that suggested by the clinical data. This result is
certainly not easy to explain to clinicians or general practitioners of
statistics. In any event, it seems worthwhile to investigate further
and see what ramifications this intriguing phenomenon may
have.\looseness=-1

To further examine the phenomenon, we compare the percentiles of the
marginal priors, the profile likelihood function and the marginal
posterior distributions of the treatment effect $\delta$ in Table~\ref{tab3}.
In each of the four Bayesian
approaches, the 95\% posterior credible interval lies inside the
corresponding 95\% interval from the prior and has substantial overlap with
the corresponding 95\% interval from the profile likelihood.
But this is not always the case at the 80\% and 90\% levels, where
several posterior credible intervals do not lie within the boundaries
of the corresponding intervals from the priors and the likelihood
functions. The outcome of whether the posterior credible interval lies
within the boundaries of the other two depends on the choice of the
credible level. Thus, using credible intervals as our primary
inferential instrument cannot completely avoid the discrepant
posterior phenomenon either.\looseness=-1

\begin{figure}

\includegraphics{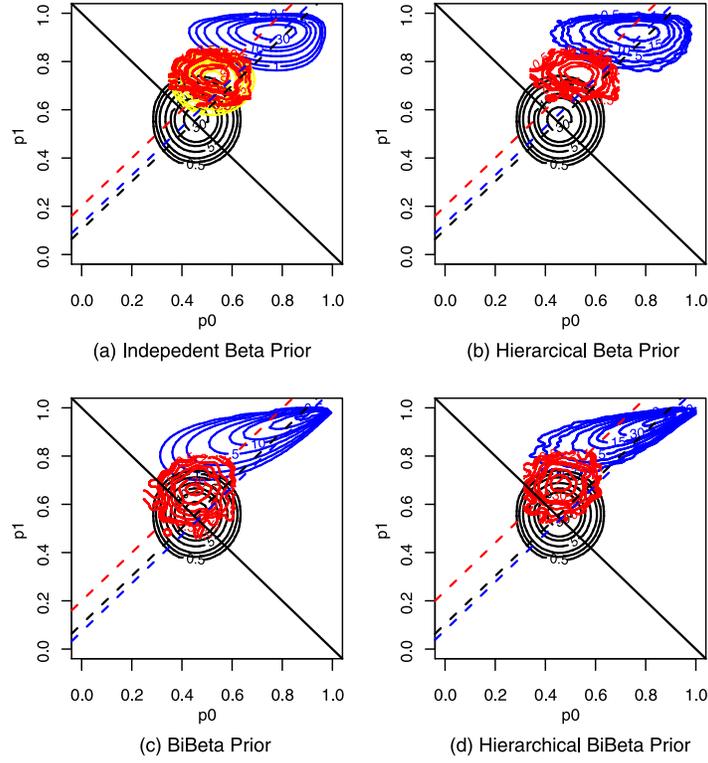}

\caption{Contour plots: joint prior distribution (in blue), joint
likelihood function (in black)
and estimated posterior distribution (in red) of $(p_0, p_1)$. These
two-dimensional distributions are
projected along the $45^o$ lines of $\delta= p_1 - p_0$ onto the
interval of
possible values of $\delta$, $-1<\delta<1$, leading to Figures
\protect\ref{fig4}(a)--(d). The yellow curves in (a) are a
posterior contour plot from the exact formula.}\label{fig5}\vspace*{-3pt}
\end{figure}

To better understand this phenomenon, we plot in Figures \ref
{fig5}(a)--(d) the
contours of the joint prior distribution $\pi(p_0, p_1)$, the
likelihood function of $(p_0, p_1)$ and the joint posterior
distribution of $(p_0,p_1)$ for each of the four full Bayesian approaches.
We show that certain projections of Figures~\ref{fig5}(a)--(d) lead to the
marginal distributions and plots in Figures~\ref{fig4}(a)--(d).
As marked in Figure~\ref{fig5}, the center (mode) of each contour plot
is on a
line $\delta=p_1-p_0$ (or $p_1=p_0+\delta$).
Varying $\delta$ in $\delta=p_1-p_0$ produces a family of parallel
lines, all making a $45^o$ angle with the horizontal axis. The
projections of the three distributions along these parallel lines onto
the interval of
possible values of $\delta$, $-1<\delta<1$, lead to the plots of
marginal distributions in Figures~\ref{fig4}(a)--(d). The
yellow curves in (a) are a posterior contour plot from
the exact formula.
Although the contour plots of the posterior distributions sit between
those of the prior distributions and the likelihood function, their
projected peaks (modes) are more to the upper-left than those of the
marginal priors and the profile likelihood function. Further
investigation indicates that this is a genuine mathematical phenomenon
which holds for all four Bayesian approaches and not merely an
aberration due to some special circumstances. In fact, when the center
(mode) of a posterior distribution is not in the interval joining the
two centers (modes) of the joint prior and likelihood functions, as is
often the case with skewed distributions (and even sometimes with
nonskewed distributions), there always exists a linear direction, say,\vadjust{\goodbreak}
$a p_0 + b p_1$ with some coefficients a and b, along which the
marginal posterior fails to fall between the marginal prior and
likelihood functions of the same parameter. Reparametrization, if done
carefully, such as considering joint distribution of $(\delta, \theta)
= (p_1 - p_0, p_1 + p_0)$ or others, may sometimes help hide the
discrepant posterior phenomenon on the $\delta$ direction, but cannot
eliminate it systematically. We have found no discussion of such a
geometric finding on marginalization in the Bayesian literature. See
further discussion in Section~\ref{sec5}.\looseness=-1

%f5 #&#
%
%f5 ###

%%%%%%%%%%%%%%%%%%%%%%%%%%%%%%%%%%%%%%%%%%%%%%%%%%%%%%%%%%%%%%%%%%%%%%%%%%%%%%%%%%%%%%%%%%%%%

%s5 #&#
%s5 ###
\section{Conclusions and additional remarks}\label{sec5}

To incorporate expert opinions in the analysis of a clinical trial with
binary outcomes in a meaningful way, we have developed and studied
several bivariate full Bayesian approaches as well as a CD approach. We
show that both the Bayesian and the proposed CD approaches may provide
viable solutions. Although the paper focuses on expert opinions in
pharmaceutical studies, the methodologies developed here\vadjust{\goodbreak} can be applied
to incorporating other types of priors or external information, for
example, historical knowledge. These methodologies should also be
useful in many other fields, including finance, social science studies
and even homeland security, where prior knowledge, expert opinions and
historical information are much valued and need to be incorporated with
observed data in an effective and justifiable manner.

In this paper we have examined and compared both Bayesian and CD
approaches. Although there does not exist the usual theoretical
platform for a direct comparison on efficiency or lengths of intervals,
the comparison can be summarized in three aspects: empirical results,
computational effort and theoretical consideration.
The empirical findings from Figure~\ref{fig2} show that, as long as the
histogram of the expert opinions can be well approximated by a normal
distribution, all approaches considered in this paper perform
comparably, in terms of the posterior distribution or the combined CD
and their corresponding inferences. However, if the histogram is
skewed, the full Bayesian approach may produce the discrepant posterior
phenomenon, which is difficult to avoid in theory and difficult to
explain in applications. The CD approach
avoids such a phenomenon.

In terms of the computational effort, the bivariate full Bayesian
approach is demanding since it requires running a large-scale
simulation using an MCMC algorithm, while the proposed CD approach is
both explicit and straightforward to compute. In addition, the CD
approach can directly incorporate the histogram of expert opinions
without an additional effort of curve fitting.

Theoretically, since it is not possible to find a ``marginal''
likelihood of $\delta$ [i.e., a conditional density function $f(\mathrm{data} |
\delta)$], any univariate Bayesian approach focusing on the parameter
of interest $\delta$ is not supported by Bayesian theory. A full
Bayesian solution is to jointly model $(p_0, p_1)$ [or a
reparameterization of the pair $(p_0, p_1)$] and, subsequently, make
inferences using the marginal posterior of $\delta$. The full Bayesian
approaches developed in the paper follow exactly this procedure and are
theoretically sound. The proposed CD approach is developed strictly
under the frequentist paradigm and is also theoretically sound. Unlike
the full Bayesian approaches, the CD approach can focus directly on the
parameter of interest $\delta$ without the additional burden of
modeling other parameters or the correlation between $p_0$ and $p_1$,
and thus appears to have some advantage in this application.

A surprising finding in this research is the discrepant posterior
phenomenon occurring in the full Bayesian approaches under skewed
priors. Although it may be mitigated if the prior is only slightly
skewed or is in accordance with the likelihood function, the phenomenon
is intrinsically mathematical. How much skewness is required to produce
the phenomenon depends on all elements involved, including shapes and
locations of both the likelihood and the prior.
The reactions to this phenomenon we have encountered thus far fall
roughly into two groups.
One group views the discrepant posterior as a mathematical\vadjust{\goodbreak} truth and,
if one has faith in the choice of the prior, one should proceed to make
inference using this marginal posterior, even though the outcome is
counterintuitive. The other worries about the counterintuitive result
and would try to find alternative approaches of a good operating
characteristic for the particular problem at hand, even at the cost of
abandoning the mathematically solid full Bayesian approach in favor of
less rigorous approaches such as the univariate Bayesian approach
described in Section~\ref{sec1}. In any case, the lesson learned from the
Bayesian analysis here is that the choice of the prior really matters
and it needs to be in some agreement with the likelihood function,
which is similar in spirit to what was referred to as ``model
dependent'' in \citet{Berg}.
We also consider this a manifestation of an inherent difficulty in
modeling accurately the joint effects of the two treatments as
reflected in $p_0$ and $p_1$ and their correlation. This difficulty
illustrates again the complexity of the practice of incorporating
external information in trials with binary outcomes.

The discrepant posterior phenomenon is caused by ``marginalization,''
but it is different from the ``marginalization paradox'' discussed in
\citet{Dawid} and \citet{Berg}. In particular, the
marginalization paradox in Dawid, Stone and Zidek (\citeyear{Dawid}) refers to the phenomenon
that the marginal posterior of $\pi(\theta| \mathrm{data})$ obtained from the
joint prior $\pi(\theta, \phi)$ and its full model $f(\mathrm{data} | \theta,
\phi)$ can sometimes be quite different (``incoherent'') from the
posterior $\pi(\theta| \mathrm{data})$ obtained by applying the Bayes formula
directly to its marginal prior $\pi(\theta)$ and marginal model $f(\mathrm{data}
| \theta)$, even though the marginal prior $\pi(\theta)$ and marginal
model $f(\mathrm{data} | \theta)$ are consistent (``coherent'') with the joint
prior $\pi(\theta, \phi)$ and the full model $f(\mathrm{data}| \theta, \phi)$.
Here, $\phi$ represents nuisance parameters. This paradox is different
from what we observed here. In our example, it is not possible to have
the marginal model $f(\mathrm{data} | \theta)$, and the discrepant posterior
phenomenon in the full Bayesian approach is that the estimate derived
from the marginal posterior $\pi(\theta| \mathrm{data})$ may not be between the
estimates from the marginal prior $\pi(\theta)$ and the profile
likelihood function $\ell(\theta| \mathrm{data})$.
This is counterintuitive in practical applications.

It is worth noting that the discussion and implications of the
discrepant posterior phenomenon extend beyond the setting of binary
outcomes to any multivariate setting involving skewed distributions. As
long as the center (mode) of a posterior distribution is not in the
interval joining the centers (modes) of the joint prior and the
likelihood function, there always exists a direction along which the
center (mode) of the marginal posterior fails to fall between the
centers (modes) of marginal prior and the profile likelihood function.
This phenomenon has implications in the general practice of Bayesian
analysis. For instance, many researchers in machine learning and other
fields routinely draw conclusions solely based on marginal posterior
distributions without checking (or it is very difficult to check) the
validity of such conclusions. The discrepant posterior phenomenon
suggests that further care is needed.\vadjust{\goodbreak}

Many methods have been introduced to model ``reported'' expert
opinions, account for their potential errors and heterogeneity, and
subsequently pool them; see \citet{Gen} for an excellent
review of this topic. In particular, Spiegelhalter, Freedman and Parmar (\citeyear{Spi94})
described ``arithmetic and logarithm pooling'' as the ``two simplest
methods'' for pooling expert opinions, and articulated a ``strong
preference'' ``for arithmetic pooling to obtain an estimated opinion of
a typical participating clinician.'' The underlying assumption of
arithmetic pooling is that the average of the ``observed'' expert
opinions is an unbiased representation of the ``true'' prior knowledge.
This assumption naturally facilitates the additive error model used in
Section~\ref{sec32} for summarizing ``reported'' expert opinions in a CD.
Clearly, the modeling principle and development we used to summarize
the expert opinions in a CD are similar in spirit to those discussed in
\citet{Gen} and Spiegelhalter, Freedman and Parmar (\citeyear{Spi94}) for Bayesian approaches.

The modeling framework developed in Section~\ref{sec32} is sufficiently
flexible and can be modified to accommodate various ways of aggregating
expert opinions. In particular, it can incorporate weighting schemes to
develop a robust method against extreme expert opinions, introduce
additional terms to reflect biased opinions or additional
uncertainties, or use the geometric mean as a way to pool the expert opinions.
Some of these extensions (e.g., the robust method) by themselves
could be attractive choices to produce priors in the context of
traditional Bayesian approaches. Due to space limitations, we will not
pursue these extensions in this paper.

In a different direction, we have also considered modeling the survey
data of expert opinions using a traditional random effects approach. In
such a model, we provide a regression model for the responses of the
100 ``virtual patients'' of each expert (as described in Section~\ref{sec11})
and add a random effect term to account for the expert-to-expert
variation. However, it seems nontrivial to overcome the technical
difficulty in making the modeling process free of the number (100) of
``virtual patients.''
In fact, this difficulty led us to the bootstrap argument in Section
\ref{sec32}, in which we mimic a potential model of expert exposure to
pre-existing experiments. Clearly, there remain many challenging issues
in modeling the survey data of expert opinions, even for the seemingly
simple binary setting.

% \section*{Appendix I: The Metropolis--Hastings Algorithm Used in
%Section 2}
%

\section*{Acknowledgments}
This is part of a collaborative research project between the Office of
Statistical Consulting at Rutgers University and Ortho-McNeil Janssen
Scientific Affairs (OMJSA), LLC.
The authors thank Dr. Karen Kafadar, the AE and two reviewers for their
constructive comments and suggestions, which have greatly helped
improve the presentation of the paper.
The authors also thank Drs. Steve Ascher, James Berger, David Draper,
Brad Efron, Xiao-li Meng, Kesar Singh and William Strawderman for their
helpful discussions and Ms. Y. Cherkas for her computing assistance on
a related MCMC algorithm.\vspace*{-3pt}

\begin{supplement}[id=suppA]\vspace*{-3pt}
\stitle{Appendix: MCMC algorithm and CD examples\\}
\slink[doi]{10.1214/12-AOAS585SUPP}  %[doi,text={...}] - jei reikia suskaldyti doi
\sdatatype{.pdf}
\sfilename{aoas585\_supp.pdf}
\sdescription{Appendix I contains a Metropolis--Hastings algorithm used
in Section~\ref{sec2}.
Appendix II presents two CD examples that are relevant to the
exposition of this paper.\vspace*{-2pt}}
\end{supplement}

% imsref loaded by akundreckaite, 2012-09-05 14:47:41
%
% imsref loaded by akundreckaite, 2012-09-06 10:06:10

\printaddresses

\end{document}